\documentclass[12pt]{article}

\usepackage[margin = 2.4cm]{geometry}
\usepackage[english]{babel}
\usepackage[utf8]{inputenc}
\usepackage{amsmath}
\usepackage{amssymb}
\usepackage{graphicx}
\usepackage{url}
\usepackage[round,sort,authoryear]{natbib}
\usepackage{rotating}
\usepackage{setspace} 
\usepackage{lineno}
\usepackage{caption}
\title{Hunting, food subsidies, and mesopredator release: the dynamics of crop-raiding baboons in a managed landscape}

\author{Rachel A. Taylor, Sadie J. Ryan, Justin S. Brashares and Leah R. Johnson}

\date{\today}

\begin{document}
\maketitle

\begin{center}

Rachel A. Taylor\footnote{rataylor@usf.edu}, Integrative Biology, University of South Florida, Tampa, Florida, 33620\\
Sadie J. Ryan, Geography and Emerging Pathogens Institute, University of Florida, Gainesville, Florida, 32611\\
Justin S. Brashares, Environmental Science, Policy and Management, UC Berkeley, Berkeley, California, 94720\\
Leah R. Johnson, Integrative Biology, University of South Florida, Tampa, Florida, 33620\\

\end{center}

\begin{abstract}
The establishment of protected areas or parks has become an important tool for wildlife conservation. However, frequent occurrences of human-wildlife conflict at the edges of these parks can undermine their conservation goals. Many African protected areas have experienced concurrent declines of apex predators alongside increases in both baboon abundance and the density of humans living near the park boundary. Baboons then take excursions outside of the park to raid crops for food, conflicting with the human population. We model the interactions of mesopredators (baboons), apex predators and shared prey in the park to analyze how four components affect the proportion of time that mesopredators choose to crop-raid: 1) the presence of apex predators; 2) nutritional quality of the crops; 3) mesopredator ``shyness'' about leaving the park; and 4) human hunting of mesopredators. We predict that the presence of apex predators in the park is the most effective method for controlling mesopredator abundance, and hence significantly reduces their impact on crops. Human hunting of mesopredators is less effective as it only occurs during crop-raiding excursions. Furthermore, making crops less attractive, for instance by planting crops further from the park boundary or farming less nutritional crops, can reduce the amount of time mesopredators crop-raid.
\end{abstract}
\vspace{5mm}
Keywords: human-wildlife conflict, mesopredator shyness, crop subsidies, crop-raiding, apex predators, trophic cascade, protected area, mathematical modeling\\ \vspace{3mm}
%% Can now have up to 10 keywords if we want to add more

\newpage

\section{Introduction}

In many African protected areas (hereafter ``parks'') the balance between humans, apex predators, mesopredators, and prey has been shifting. In most parks apex predators (e.g.\ lion, leopard, spotted hyena) are disappearing due to poaching \citep{brashares2004bushmeat, kaltenborn2005nature}, disease \citep{murray1999infectious}, and habitat conversion or reduction \citep{balme2010edge}. Simultaneously, the density of human populations around parks is increasing \citep{wittemyer2008accelerated, joppa2009population}, compounding pressure on apex predators and increasing mortality due to poaching and human-wildlife conflict \citep{lindsey2005attitudes,woodroffe2007livestock}. In many parks their prey, such as ungulates, are also subject to high rates of hunting for meat \citep{thirgood2004can}. While apex predators and their prey are decreasing, in contrast populations of mesopredators, such as baboons, are increasing. For example, over the period 1968 -- 2004 all large apex predators became extinct in 3 of the 6 parks studied by the Ghana Wildlife Division, but baboons had a 365\% increase in observations and 500\% increase in range \citep{brashares2010ecological}. This increased abundance of baboons, which occurs in many parks throughout Africa, frequently results in crop-raiding as the populations spill over into farmed land outside the park \citep{hill1997crop, hill2000conflict}. Baboons are able to exploit all trophic niches. They can act as predators, compete for browse with ungulates and livestock \citep{strum1982variations}, and exploit domesticated landscapes. It has been shown that baboons respond quickly to newly available resources in terms of fecundity \citep{bercovitch1993dominance}, and have a high rate of potential demographic increase for a large primate. Given sufficient resources baboon populations could potentially increase at roughly 20\% annually (see Appendix A).

Increased baboon abundance can pose a serious problem for people living at or near the boundaries of parks because of the potential for a large percentage and wide variety of crops to be destroyed, even during single crop-raiding events \citep{naughton1998predicting,naughton1997farming,tweheyo2005patterns,hartter2014contrasting}. This destruction of crops is detrimental to human livelihood and education (due to children and women staying in the fields to defend the crops \citep{mackenziechasing}). It can also contribute to negative attitudes toward the park, potentially undermining conservation aims \citep{tweheyo2005patterns, hartter2014contrasting,ryan2015household}. Thus, understanding why baboons crop-raid, and, importantly, how to control levels of crop-raiding, could be very beneficial both to park conservation goals and the livelihoods of people living near them.

In a recent study, \cite{Nishijima2014roles} presented a mathematical model of mesopredator release when the mesopredator has both prey shared by an apex predator and alternative, unshared prey. In particular, they examined the effect of including alternative prey on the change in shared prey abundance when apex predators are lost. Although mesopredator release does not necessarily have a negative effect on the shared prey species \citep{brashares2010ecological}, a large supply of alternative prey can intensify mesopredator release and its deleterious effects on the ecosystem.  

We expand on the framework developed by \citet{Nishijima2014roles} to understand the influence of human subsidies on this crop-raiding mesopredator dynamic. In this case the subsidy is crops, which exist outside the boundaries of parks. The continual availability of crops could enhance baboon population increases and thus feed back into more crop-raiding. However, unlike the unlimited alternative resource of \cite{Nishijima2014roles}, in this model we assume that the boundaries of the park pose an added consideration for baboons. The presence of humans, domestic animals, and reduced cover may make baboons reticent about leaving the park, even while increased baboon abundance within the park increases pressure to leave the park to crop-raid. By incorporating a more complete mathematical description of this behavioral trade-off in potential baboon crop-raiding, we can link previous work on mesopredator release with human subsidy control and management strategies, to gain an understanding of the balance in managing the damaging effects of crop-raiding. 

In African parks, many animals crop-raid including multiple primate species (e.g. gorillas \citep{hill1997crop,biryahwaho2002community}; chimpanzees
\citep{madden1999management}; redtail, colobus, vervet, blue \citep{hill1997crop} and golden \citep{biryahwaho2002community} monkeys), ungulates such as buffalo, bushbuck, and duikers (e.g. \cite{plumptre1997effects}), and elephants (e.g. \cite{naughton2005socio, osborn2004seasonal}). 
These animals may also face behavioral barriers to leaving the park to crop-raid, and will in many cases be affected by predation and/or human hunting. Therefore, our analysis can extend to a broad range of crop-raiding species rather than focused purely on baboons. Additionally, mesopredator release with human subsidies is not limited to parks landscapes in Africa. For example, dingo abundance in Australia is carefully controlled due to their predation upon farmed cattle. However, this has led to mesopredator release of foxes and cats and concurrent declines of their prey; foxes often benefit from human subsidies as sheep farming leads to an abundance of rabbits \citep{johnson2007rarity}. Whilst this is not a direct analogue to our system, parallels can be drawn to give insight into a wider spectrum of ecosystems. We thus present a fairly specific case to illustrate a framework that is flexible to many scenarios.

In this paper we use a mathematical model to explore how predation, human hunting, mesopredator shyness, and quality of crops can all interact, leading to different proportions of time spent crop-raiding by a mesopredator. Model parameters are chosen with an eye towards the particular case of baboons, although the model is more general. Our model could be adapted to any scenario in which there are two competing species with a shared predator, where one of the competitors has access to a separate source of food such as human subisides \citep{oro2013ecological}. We explore the sensitivity of the dynamics to the parameters (including the willingness of the baboons to leave the park), and to changes in three potential controls: hunting of apex predators, hunting of mesopredators, and quality of the food subsidies. We use this analysis to assess how crop-raiding may be reduced by human or apex predator control.

\section{Model}

\subsection{Model Outline}

\begin{figure}
\centering
\includegraphics[width = 13cm]{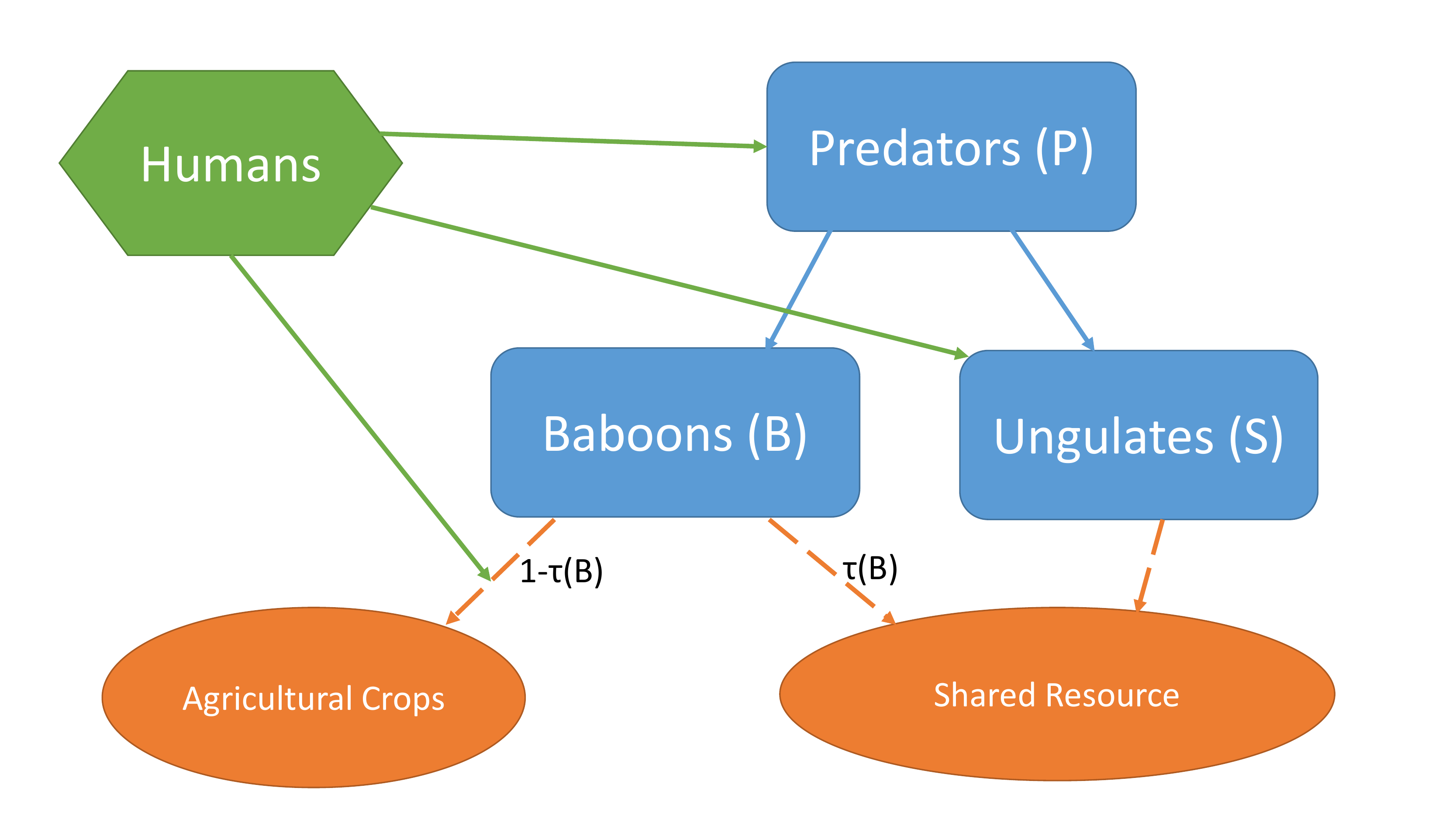}
\caption{The interactions between the apex predators, baboons and ungulates with their available resources and the role of human hunting. Variables directly included in the model are shown as blue rectangles, orange ovals represent resources which impact carrying capacities, and the green hexagon shows the impact of human hunting. Baboons spend a proportion of their time, $\tau(B)$ living off the park resources and the rest of their time, $1 - \tau(B)$, crop-raiding. Baboons are only hunted by humans while crop-raiding, and baboons and ungulates share an in-park resource pool, which determines carrying capacity.}
\label{fig:schematic}
\end{figure}

We model the population abundances of mesopredators (baboons), apex predators, and shared prey (ungulates) as they interact with each other, their shared resources, and with human hunting. We use the term ungulates to represent many potential species of ungulates within the park. A schematic for the modeled ecosystem is provided in Figure \ref{fig:schematic}. This shows that apex predators, at the top of the food-chain, prey upon both the baboons and the ungulates. The park itself provides resources for the baboons and ungulates, both in terms of food availability and living space. We do not model this resource explicitly, but rather implicitly through the carrying capacity of baboons and ungulates, i.e.\ how many of these animals the park is able to sustain. The baboons and ungulates compete for this shared resource; their impact on each other arises from how well they exploit the shared resource. 

Baboons have an additional resource of agricultural crops, inaccessible to ungulates, which increases the number of baboons that can be sustained within this system as they are only restricted by spatial and social requirements \citep{warren2011crop}. Furthermore, agricultural crops are often high in nutritional value therefore baboons are able to reproduce at a higher frequency when they feed off crops compared to the park resources. 

The human population outside the park is also implicit. We need only consider its impact on the apex predators, baboons, and ungulates. In this model humans hunt all three animal groups. However, they only hunt baboons because they raid agricultural crops, thus hunting only occurs when the baboons are crop-raiding. 

The model assumes that baboons spend a portion of their time, $\tau (B)$, living off the resources of the park, in competition with ungulates, and the rest of their time crop-raiding, $1-\tau (B)$. We assume that crops are available for baboons through the entire year. We translate this into a mathematical model (Eqns.~\eqref{eqn:full1}-\eqref{eqn:full}) where $B(t)$, $S(t)$ and $P(t)$ are the numbers of baboons, ungulates and apex predators at time $t$ respectively: 

\begin{align}
\frac{dB}{dt} = & r_B B \left \{ \left (1 - \frac{B}{K_B} - \frac{a_{BS} S}{K_B} \right) \tau (B) + \alpha \left (1 - \tau (B) \right) \left( 1 - \frac{B}{\beta K_B}\right) \right \} \label{eqn:full1}\\ & - H_B (1 - \tau (B)) B - \delta_P f_{BP} (B)P \nonumber\\
\frac{dS}{dt} = & r_S S \left \{ \left (1 - \frac{S}{K_S} \right ) - \frac{a_{SB} B}{K_S}\tau (B)  \right \} - H_S S - (1- \delta_P) f_{SP} (S)P \label{eqn:full2} \\
\frac{dP}{dt} = & \epsilon_B\delta_P f_{BP} (B)P + \eta \epsilon_B(1- \delta_P) f_{SP} (S)P - (\mu + H_P)P.
\label{eqn:full}
\end{align}
Both baboon and ungulate populations experience logistic growth with growth rates ($r_B, r_S$) and carrying capacities ($K_B, K_S$) determined by the available resources in the park. However, this is regulated by competition, with $a_{BS}$ being the rate of competition by $S$ on $B$. The baboons only spend a proportion of their time $\tau (B)$ feeding on park resources and competition between baboons and ungulates only occurs during this time. Hence, the competition term in \eqref{eqn:full2} is multiplied by $\tau (B)$. When baboons are feeding on agricultural crops the population growth is still determined by a logistic term, but their growth rate is increased by rate $\alpha$ and the carrying capacity by rate $\beta$ to represent the benefits of crop-raiding -- enhanced nutritional quality of food and reduced limitations on availability. All three populations are affected by human hunting ($H_B$, $H_S$, $H_P$), but baboons only when they are crop-raiding, hence  $H_B$ is multiplied by $1-\tau (B)$ in \eqref{eqn:full1}. Apex predators survive and reproduce by feeding upon both baboons and ungulates, spending a proportion $\delta_P$ of their time hunting baboons and the rest of their time hunting ungulates. Apex predators are also free to roam outside of the park and hence they are able to hunt baboons at all times regardless of whether baboons are crop-raiding (the predation terms do not contain $\tau(B)$). For now we use a general functional form ($f_{BP}(B), f_{SP} (S)$) to represent the predation terms, which could change for different ecosystems. For a full description of the parameters and their values, see Table \ref{tab:par}.

There is some evidence that baboons prey upon ungulates opportunistically, such as Thomson's gazelles in Gilgil, Kenya \citep{strum1982variations, bercovitch1993dominance} and on Kob and goats, in and near parks in Ghana ({J. Brashares Pers. Obs.}). This conforms to a mesopredator release scenario in that loss of the top apex predators allows baboons to become the mesopredator, controlling ungulate population abundance both through competition for resources and by predation. We model this indirectly by imposing a large competitive effect of baboons on ungulates. In our model the ungulate population represents a number of ungulate species, as well as potentially smaller primates, all of whom are in competition with the baboons for park resources. Since baboons will not prey upon all species of ungulates we do not model direct baboon predation upon this category but rather subsume it into competition.

\begin{sidewaystable}[!p]
\small
\begin{tabular}{c c c}
\hline
Parameter & Interpretation & Value \\ \hline
$r_B$ & Growth rate of baboons when feeding upon park resources & 0.2 \\
$r_S$ & Growth rate of ungulates &0.35\\
$K_B$ & Carrying capacity of baboons from park food resources & 1300\\
$K_S$ & Carrying capacity of ungulates from park food resources & 6000\\
$a_{BS}$ & Competition effect of S (ungulates) on B (baboons) & 0.1\\
$a_{SB}$ & Competition effect of B (baboons) on S (ungulates) & 1.1\\
$\alpha$ & Proportional effect of crop-raiding on growth rate of baboons & 1.49\\
$\beta$ & Change in baboon carrying capacity limited by space/social confines only & 1.27\\
$\sigma$ & Strength of switching to crop-raiding & 3\\
$\gamma$ & Effect of crowding (baboon shyness) & 0.5\\ 
$H_B$ & Hunting rate of baboons by humans & 0\\
$H_S$ & Hunting rate of ungulates by humans & 0.01 \\
$H_P$ & Hunting rate of apex predators by humans & 0\\
$f_{BP}(B) = \frac{m_B B}{B + G_B}$ & Functional form for predation of baboons - max predation rate & $m_B = 45$\\
& Half saturation constant &  $G_B = 800$\\
$f_{SP}(S) = \tfrac{m_S S}{S + G_S}$ & Functional form for predation of ungulates - max predation rate & $m_S = 80$\\
& Half saturation constant & $G_S = 2500$\\
$\epsilon_B$ & Efficiency conversion of baboon predation into growth of apex predators & 0.005 \\ 
$\eta$ & Change in efficiency conversion of ungulate predation compared to baboon & 1.3\\ 
$\mu$ & Natural death rate of apex predators & 0.06\\\hline
$\tau(B)$ & Proportion of time baboons feed upon park resources & -\\
$\delta_P$ & Proportion of time apex predators hunt baboons & - \\ \hline
\end{tabular}
\caption{Parameter descriptions and values for the model described in Equations \eqref{eqn:full1}-\eqref{eqn:full}. Rates are per year. References for the parameter values can be found in \S \ref{sec:par}.}
\label{tab:par}
\end{sidewaystable}

\subsection{Time Spent Crop-Raiding}

To understand the proportion of time that baboons focus on park resources, $\tau (B)$, we have to make some assumptions about the factors determining when baboons crop-raid. We assume baboons choose optimally how to split their time between crop-raiding and park resources, based upon the benefits and costs of each action. Built into this is the idea that baboons will always spend some portion of their time feeding off park resources. The baboon population can grow with rate $r_B$ when utilizing park resources and at rate $r_B \alpha$ when crop-raiding. However, they are susceptible to human hunting when crop-raiding at rate $H_B$. If we did not include any impact of shyness, previous studies on patch choice \citep{krivan1997dynamic} indicate that baboons would optimally split their time to feed proportionally upon park resources and upon crops in the ratio $\tfrac{r_B}{r_B + r_B\alpha - H_B} : \tfrac{r_B\alpha - H_B}{r_B + r_B\alpha - H_B}$ respectively. However, we additionally consider that their level of crop-raiding is based upon a crowding parameter $\gamma$ (see Table \ref{tab:par}): baboons are "shy" about leaving the park, but higher baboon numbers encourages them to do so. This is introduced by considering the number of baboons in relation to their carrying capacity and how shy they are, i.e.\ $\tfrac{B}{K_B} - \gamma$. This leads to the following form for $\tau (B)$:
\begin{equation}
\tau (B) = \frac{r_B}{r_B + r_B\alpha - H_B} + \left(\frac{r_B\alpha - H_B}{r_B + r_B\alpha - H_B}\right)\left( \frac{1}{1 + \exp(\sigma(\tfrac{B}{K_B} - \gamma))}\right).
\label{eqn:tauh}
\end{equation}
The exponential term allows a smooth transition from 100\% park resources to a combination of resources and crop-raiding, as $B$ increases. The first term in this equation is the minimum proportion of time that baboons will spend feeding upon park resources, which is supplemented by the second term when either shyness is high or baboon numbers are low. As B increases, the last term will approach 0 which indicates that baboons are spending the minimum proportion of time only in the park. Hence, the proportion of time crop-raiding would approach $\tfrac{r_B\alpha - H_B}{r_B + r_B\alpha - H_B}$.

Similarly, we assume that apex predators split their time predating upon baboons and ungulates based upon the relative benefits from each source, which leads to
\begin{equation}
\delta_P = \frac{\epsilon_B f_{BP} (B)}{\epsilon_B f_{BP} (B) + \eta\epsilon_B f_{SP} (S)} = \frac{f_{BP} (B)}{f_{BP} (B) + \eta f_{SP} (S)}
\end{equation}
for the proportion of time the apex predators spend predating baboons (and $1-\delta_P$ for ungulates). $\epsilon_B$ and $\eta\epsilon_B$ are the rates determining how efficient the apex predators are at converting prey (baboons and ungulates, respectively) into reproduction. The functional forms for the predation of baboons and ungulates by apex predators are taken to be of Holling Type II form, such that 
\begin{equation}
f_{BP}(B) = \frac{m_B B}{B + G_B},
\end{equation}
where $m_B$ is the maximum predation rate on baboons and $G_B$ is the number of baboons at which predation is half its maximum. A similar equation holds for $f_{SP}(S)$ (see Table \ref{tab:par}).

\subsection{Parameter Estimates}
\label{sec:par}

Parameters are estimated from literature where possible, with the parameters calibrated to produce dynamically realistic scenarios for co-existence of baboons, ungulates and apex predators (Table \ref{tab:par}). We use available data from Greater Addo Elephant National Park, South Africa, which includes forested parkland but scale it to a smaller, more average park size. Since we assume that the $S(t)$ class can represent many species of ungulates, and even small primates, with which the baboons compete for park resources, we add together estimates of ungulate abundance for the carrying capacity and use average birth and death rates from multiple species to calculate the growth rate. Growth rates for baboons are found in \cite{smuts1989reproduction} and \cite{Ryan2015Data} and carrying capacities in \cite{hayward2007carrying}, with the higher estimates used to quantify the increases due to crop-raiding. Growth rates for ungulates are found in \cite{spinage1972african} and carrying capacities in \cite{hayward2007carrying}. Estimates for apex predator longevity are found in \cite{paemelaere2011fast} as well as information on fecundity rates to inform parameters $\epsilon_B$ and $\eta$.  Estimates for the maximum predation rates ($m_B$, $m_S$) and half-saturation constants ($G_B$, $G_S$) were taken from \cite{altman1968metabolism}. Average population size of apex predators within different parks is found in \cite{hayward2007carrying}, which was used to confirm the population estimates were reasonably accurate. The parameters chosen are intended for a conceptual understanding of how the different species interactions affect baboon crop-raiding generally. A more in-depth analysis at a specific location would require species data at that location.

\subsection{Method Outline}

We are interested in the effect of varying the parameters on the amount of time the mesopredator is expected to crop-raid. We initially take a local stability approach, whereby we numerically find the solution to the model (Equations \eqref{eqn:full1}-\eqref{eqn:full}) through simulations while varying one or more parameters and keeping the others constant at the values given in Table \ref{tab:par}. We then calculate the average time spent crop-raiding once the solution has reached a steady state or cyclic solution. Cyclic solutions occur when baboons, ungulates and apex predators co-exist and, thus, to calculate the average time spent crop-raiding over the cyclic solution the simulation is run for 500 years and the average taken over the last 100 years. Whilst the long time scale is unrealistic for conservation strategies it is required to allow the transient dynamics to fade. Further, primates and apex predators are long-lived animals which leads to long period cycles so a long time scale is necessary to calculate the mean proportion of time spent crop-raiding. 

We use two different measures to determine crop-raiding effects, both $1-\tau(B)$ and $(1-\tau(B)) B$. While only slightly different, they provide two ways to view crop-raiding dynamics -- from the baboon perspective and from the human perspective. $1-\tau(B)$ is the proportion of time a single baboon will spend crop-raiding, hence it informs us of the level of enticement for baboons to crop-raid. $(1-\tau(B)) B$ informs us of the impact the whole population of baboons, crop-raiding at that level, will have on humans. It can be thought of as the number of baboons that choose to crop-raid 100\% of their time, while the rest of the baboons do not crop-raid at all. This perhaps reflects reality in that some troops of baboons do not crop-raid at all and other troops may solely subsist by crop-raiding. Hereafter, $\tau(B)$ will be shortened to $\tau$.

\section{Results}

%We are interested in how the level of crop-raiding depends on four factors and their interactions: (1) the population of apex predators, determined by the level of human hunting ($H_P$); (2) the nutritional quality of the crops ($\alpha$); (3) baboon shyness ($\gamma$); and (4) human hunting of baboons ($H_B$). Thus we will focus on these parameters and their effects on $1-\tau$ (the proportion of time baboons spend crop raiding) to understand in more detail the potential for baboons to crop-raid and whether apex predator or human control is more efficient at reducing it.

We will consider whether apex predator or human control is more efficient at reducing the proportion of time baboons spend crop-raiding, ($1-\tau$), by focussing on four factors and their interactions: (1) the population of apex predators, determined by the level of human hunting ($H_P$); (2) the nutritional quality of the crops ($\alpha$); (3) baboon shyness ($\gamma$); and (4) human hunting of baboons ($H_B$).

\subsection{Effect of apex predator removal on baboons and crop-raiding levels}

\begin{figure}[ht!]
\begin{centering}
\includegraphics{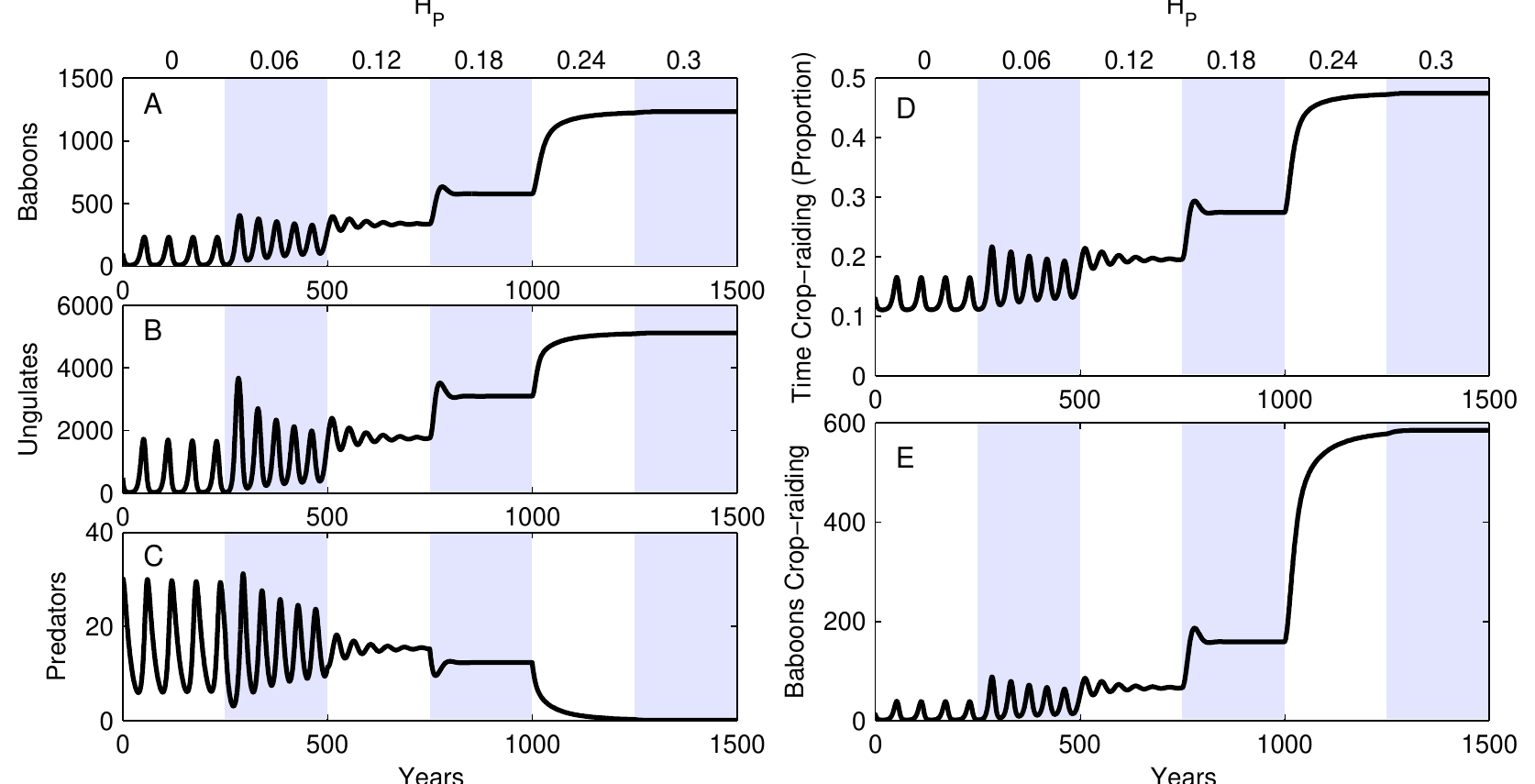}
\end{centering}
\caption{Gradual increase of hunting on the apex predators ($H_P$) to show the potential of mesopredator release with loss of apex predators. In $A$, $B$ and $C$ the dynamics of baboons, ungulates and apex predators over time, respectively. In $D$ the proportion of time baboons choose to crop-raid, ($1-\tau$), as well as the potential damage caused when they do so in $E$, measured as the number of baboons raiding, $(1-\tau)B$, are plotted against time. Parameter values are as in Table \ref{tab:par} apart from $H_P$ which varies from 0 to 0.3 every 250 years in increments of 0.06, as indicated on the top axis. The long time scale is used to show the cyclic dynamics at each level of hunting in more detail.}
\label{fig:GradPredHunt}
\end{figure}

We first consider the potential for mesopredator release when apex predators are lost through overhunting, but hunting of baboons does not occur. In Figure \ref{fig:GradPredHunt}(A,B,C) we show the dynamics of the system as hunting of apex predators is increased over time, eventually decreasing the apex predators to the point of extinction. This shows the striking increases in both baboon and ungulate populations that occur due to the loss of the apex predators. For example, the baboon population starts at an average of 200 baboons but increases to over 1000 as apex predators are driven to extinction.  

Removing apex predators, and the resulting increase in the baboon population size, has a knock-on effect for crop-raiding. In Figure \ref{fig:GradPredHunt}D we show the proportion of time that baboons choose to crop-raid as the apex predator population is reduced. Due to the crowding parameter ($\gamma$), the baboons only crop-raid when they have higher population abundances, which results in baboons spending nearly half their time crop-raiding when apex predators are absent. However, the change in the effect of crop-raiding in terms of the number of baboons leaving the park, $(1-\tau)B$, is even more dramatic, growing tenfold due to the loss of apex predators (Figure \ref{fig:GradPredHunt}E). When human hunting of baboons is present the graphs show a similar shape and pattern, but increases in population abundance of baboons and the proportion of time spent crop-raiding are suppressed to lower levels (see Appendix B).

\subsection{Effectiveness of human efforts to control crop-raiding}

\subsubsection{Hunting of mesopredators and crop quality}

\begin{figure}[ht1!]
\centering
\includegraphics{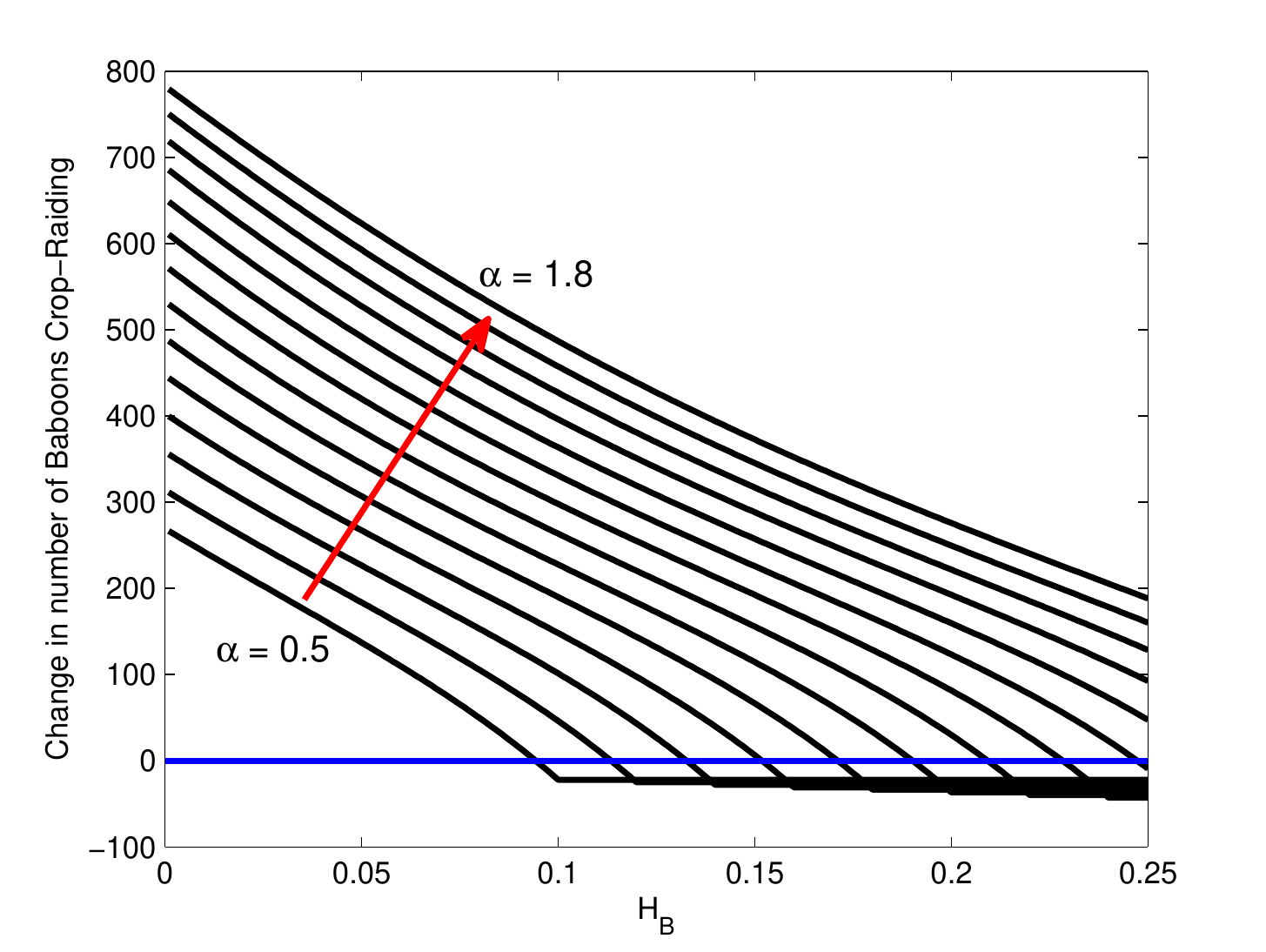}
\caption{Varying the nutrition of the crops ($\alpha$) can lead to differences in the impact of the crop-raiding. The increase in number of baboons crop-raiding, $(1-\tau)B$, from when they are controlled purely by apex predator presence ($H_P = 0$, $H_B=0$) to purely human hunting ($H_P = 0.3$, apex predators are extirpated) is shown against different levels of human hunting ($H_B$) for different values of $\alpha$, the crop-induced increase in the growth rate. $\alpha$ increases from 0.5 to 1.8 in increments of 0.1, as indicated by the arrow. All other parameters are as in Table \ref{tab:par}. The blue line indicates when there is no net increase or decrease in baboons crop-raiding.}
\label{fig:diffalp}
\end{figure}

The dynamics shown in Figure \ref{fig:GradPredHunt} assume that human hunting of baboons to discourage crop-raiding is not occurring. We now consider whether humans are able to replicate the effect of apex predators by hunting baboons during crop-raiding events. We examine human imposed baboon control both in the presence and absence of apex predators to see how much hunting of baboons is required to keep the level of crop-raiding low, and how it depends on apex predators. However, the energetic value of crops is another aspect affecting the propensity of the baboons to crop-raid. Higher quality crops will increase the growth rate of baboons more than lower quality crops, increasing the attractiveness of crop-raiding, and increasing the probability that baboons will raid even if they are generally shy. To understand in detail the amount of human hunting of baboons which is required to discourage the crop-raiding events, we consider how different levels of crop attractiveness affects the number of baboons crop-raiding. In  Figure \ref{fig:diffalp} we show the difference in the number of baboons crop raiding, $(1-\tau)B$, between two scenarios: apex predators present and no human hunting of baboons occurs versus apex predators absent and human hunting is used to control crop-raiding. We calculate this difference for a range of values of crop quality, $\alpha$. We are able to directly see the change in effectiveness of the two different control strategies working separately. Positive values in Figure \ref{fig:diffalp} indicate that the number of baboons crop-raiding increases by this amount after apex predators are extirpated and human hunting occurs instead. For example, if we take the lowest black line, $\alpha = 0.5$, and trace along this as $H_B$ increases, it shows that replacing apex predator control with human hunting at the rate $H_B$ will lead to increases in the number of baboons crop-raiding until $H_B = 0.1$.

Figure \ref{fig:diffalp} shows that virtually all values of $\alpha$ and $H_B$ lead to increases in the number of baboons crop-raiding when subject to human rather than apex predator control, with increases in $\alpha$ more than doubling the number of baboons crop-raiding at low hunting levels. Even for high levels of hunting of baboons (e.g.\ $H_B = 0.25$), when crops are more nutritional ($\alpha > 1.4$), there will always be an increase in number of baboons choosing to crop-raid 100\% of the time when subject to human hunting rather than apex predator control. The horizontal lines for lower $\alpha$ occur because the baboons spend all their time in the park once human hunting is introduced (and predators are absent) as human hunting and the lack of good resources from crops make crop-raiding no longer worthwhile. Thus, if the crops are of poor nutritional value for baboons and human hunting is high, baboons will completely stop crop-raiding. In this case, control by apex predators leads to minimally more crop-raiding (negative values in Figure \ref{fig:diffalp}) than when apex predators are lost and human hunting is used as a control. However, both high levels of human hunting and poor nutritional crops ($\alpha < 1$ indicates that the crops are less nutritional than the resources of the park) must be present for this to occur. Yet the baboons are predominantly still choosing to crop-raid unless hunting reaches high values; for example, for $\alpha = 1$, $H_B$ needs to be above approximately 0.19 to deter more crop-raiding under hunting than predation. This highlights the strength of apex predator over human control in its effectiveness of reducing crop-raiding.

\subsubsection{Mesopredator shyness and crop-raiding propensity}

\begin{figure}[ht1!]
\centering
\includegraphics{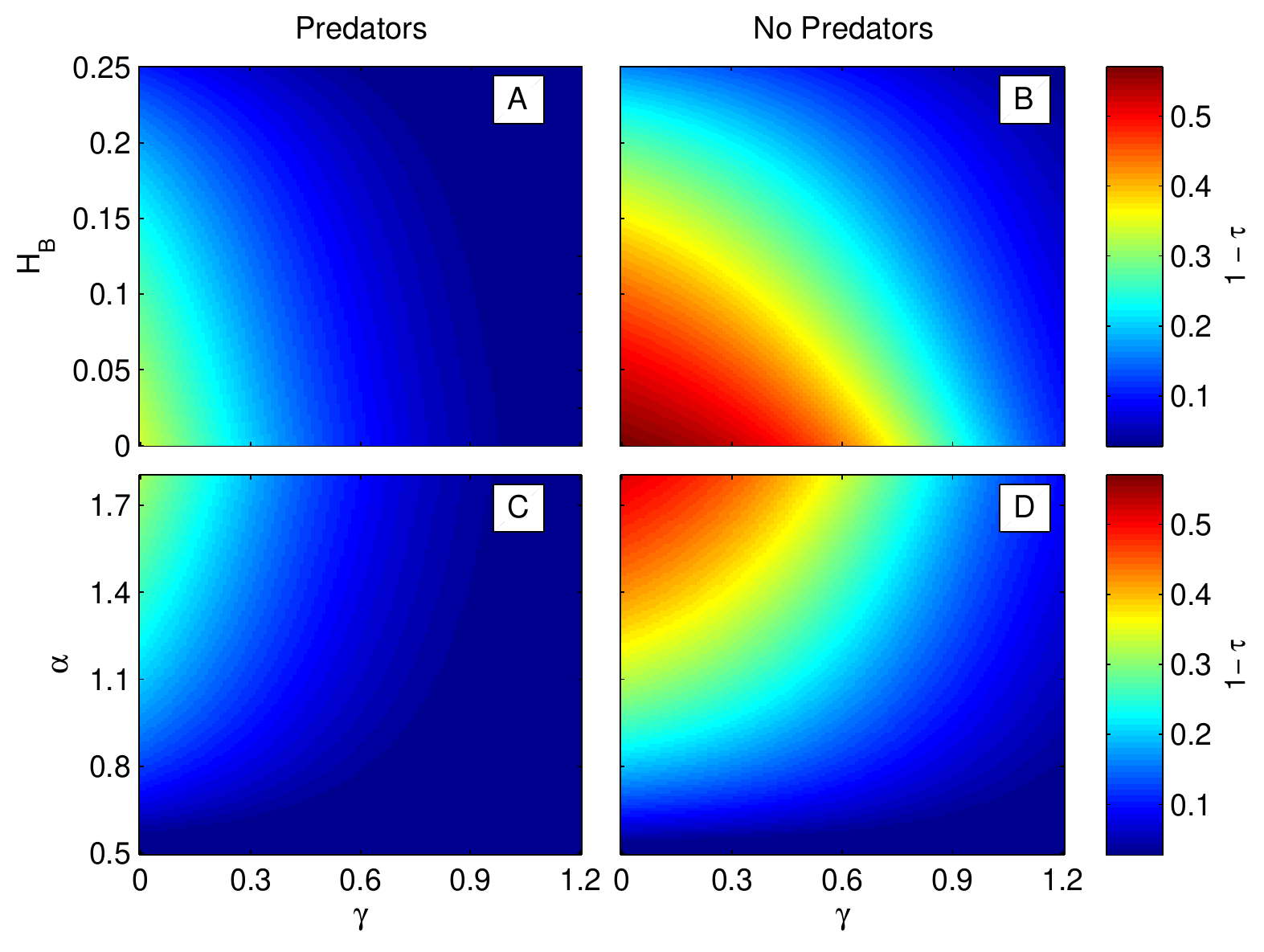}
\caption{The proportion of time baboons crop-raid ($1-\tau$) depending on baboon shyness ($\gamma$) about exiting the park. Baboon shyness is plotted against human hunting ($H_B$) in $A,B$ and against the nutritional value of crops ($\alpha$) in $C,D$. In $A,C$ apex predators are present in the park ($H_P=0$); in $B,D$ apex predators are absent in the park ($H_P=0.3$). In $A,B$, $\alpha = 1.49$ and in $C,D$, $H_B = 0.1$. All other parameters are as in Table \ref{tab:par}.}
\label{fig:alpvsgam}
\label{fig:huntalpvsgam}
\end{figure}

Shyness of the mesopredator will also affect their crop-raiding propensity. We analyze the potential for baboon shyness to affect crop-raiding time in Figure \ref{fig:huntalpvsgam}, both for apex predators present (Figure \ref{fig:huntalpvsgam}A,C) and apex predators absent (Figure \ref{fig:huntalpvsgam}B,D). We consider the impact of baboon shyness alongside two different management strategies, human hunting in Figure \ref{fig:huntalpvsgam}(A,B) and planting crops of different nutritional value in Figure \ref{fig:huntalpvsgam}(C,D). 

As expected, lower $\gamma$, i.e.\ reduced shyness, and lower hunting levels induce more crop-raiding both for apex predators present and apex predators absent (Figure \ref{fig:huntalpvsgam}A,B). When shyness is low crop-raiding will occur regardless of how much hunting pressure is applied. However, in the case when apex predators are absent (and populations are larger) baboons are more likely to crop-raid, including for higher shyness levels. The suppression of baboons by predators is amplified by their shyness. Shy baboons will only leave the park when they are forced to, due to high abundance and crowding, which corresponds to low apex predator numbers. The more gradual change in crop-raiding in the horizontal direction for low $H_B$ in Figure \ref{fig:huntalpvsgam}B compared to Figure \ref{fig:huntalpvsgam}A highlights the fact that the increased density of baboons within the park requires $\gamma$ to be very high in order to effectively reduce crop-raiding.

In Figure \ref{fig:huntalpvsgam}(C,D) we focus on the role of the nutritional value of crops ($\alpha$) and mesopredator shyness ($\gamma$), for a low level of human hunting ($H_B = 0.1$). In Figure \ref{fig:huntalpvsgam}D, even with high $\gamma$,  and human hunting occurring, the absence of apex predators leads to baboons still choosing to crop-raid for a small proportion of time if the crops are beneficial enough. Again the presence of apex predators reduces the amount of time the baboons spend crop-raiding. However, due to human hunting of baboons, not all crops will coax baboons out of the park.

It is possible to compare in Figure \ref{fig:huntalpvsgam} the two strategies of human hunting and lower nutritional crops for different values of baboon shyness. When $H_B = 0$ there are higher levels of baboon crop-raiding than when $\alpha = 1.8$, indicating that keeping the nutritional value of crops high is of less detriment than reducing the amount of human hunting of baboons. This is true for all values of baboon shyness.

\subsection{Mixed hunting strategy}

\begin{figure}[!ht]
\includegraphics{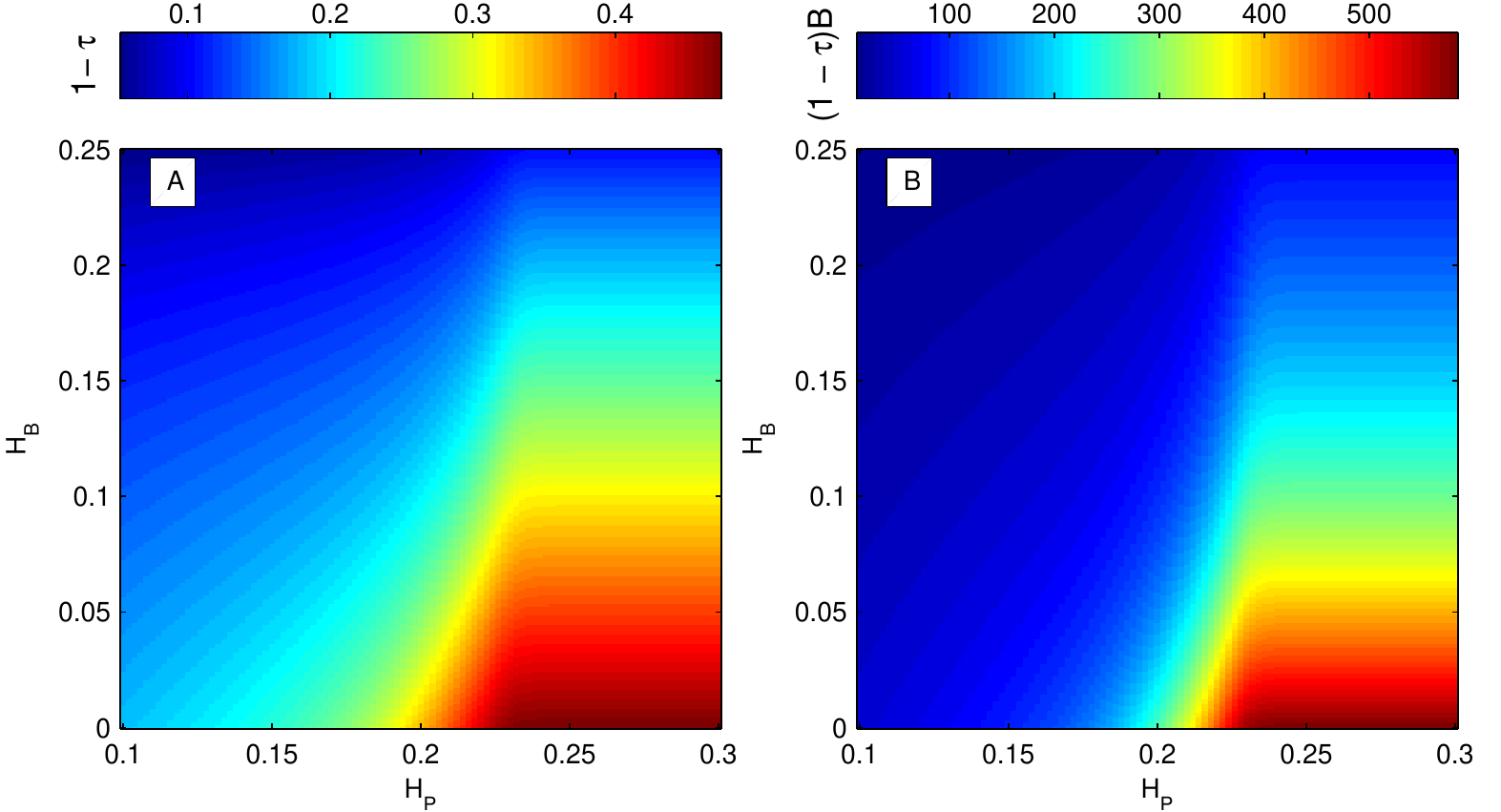}
\caption{The effect of human hunting of baboons ($H_B$) and abundance of apex predators ($H_P$) on crop-raiding. In $A$ the proportion of time spent crop-raiding ($1-\tau$) is shown against $H_B$ and $H_P$. In $B$ the number of baboons crop-raiding, i.e.\ $(1-\tau)B$ is shown against $H_B$ and $H_P$. All other parameters are as in Table \ref{tab:par}.} 
\label{fig:huntvspred}
\end{figure}

In the previous figures we focused on the two extreme cases of no apex predator hunting by humans or a complete absence of apex predators (for example, from over-hunting). Now we wish to consider the effect of different mean abundances of predators due to changes in hunting impact. We examine the effects of intermediate levels of apex predator hunting without extirpation on crop-raiding in Figure \ref{fig:huntvspred}. 

As $H_P$ increases beyond approximately 0.23 in Figure \ref{fig:huntvspred} the level of hunting on apex predators no longer affects the proportion of time spent crop-raiding, rather it is changing due to $H_B$ only. This is because apex predators die out for higher values of $H_P$. As $H_P$ approaches 0.23, the amount of crop-raiding increases. A sharper increase occurs for the number of baboons crop-raiding -- in Figure \ref{fig:huntvspred}$B$ the color scale changes more quickly as $H_P$ approaches 0.23. This is because both the time spent crop-raiding and the actual number of baboons increase due to loss of apex predator suppression, hence $(1-\tau)B$ increases at an even faster rate. This figure highlights the potential to have control over apex predator abundance from hunting without reducing the benefits of apex predator suppression of the baboon population, since for $H_P < 0.19$ there are very low levels of crop-raiding occurring. This is true even when human hunting of baboons is at low levels. For further details showing the effects of this mixed hunting strategy on apex predator abundance, see Appendix B.

\section{Discussion}

In this study we used a mathematical model to examine the trade-offs inherent in mesopredator release due to apex predator extirpation. We focused on an application to crop-raiding baboons in African parks, and the potential for crop-raiding to be mitigated by human control via hunting of baboons, predator population maintenance, or crop choice.

Our results show that the presence and abundance of apex predators is one of the most effective strategies for reducing time spent crop-raiding by baboons as the apex predators suppress baboon abundance. Even when the crops are of high nutritional value, baboons exhibit little shyness, and no human hunting of baboons occurs, the proportion of time baboons crop-raid remains low as long as apex predators are present in the park to control population numbers. We also predict that it is possible to control both apex predator and baboon populations at the same time, while maintaining low levels of crop-raiding. As long as hunting of apex predators is kept below a certain level, apex predator numbers can be controlled and little hunting of baboons is necessary to keep crop-raiding at low levels.  People living near parks may be unhappy with a high abundance of apex predators present in the park even if they control baboon crop-raiding, or park managers may wish to keep apex predator abundance under control as part of conservation strategies. Thus, a successful approach could be to combine the management strategies of hunting of apex predators and of baboons. 

Human hunting of baboons on its own is not able to replicate the suppression of baboons that occurs under predation, and so is not typically able to significantly lower the time spent crop-raiding in comparison. This is especially true if crops have high nutritional value and so induce the baboons to crop-raid. Figure \ref{fig:diffalp} highlights the potential for reducing the number of baboons crop-raiding by lowering $\alpha$, the nutritional value of the crops; it is possible to halve the number of baboons crop-raiding when $\alpha$ is reduced. One potential management strategy could be to replace crops such as maize by less nutritious crops, such as tea.

Higher baboon shyness leads to a reduction in the time spent crop-raiding by baboons. Unfortunately, baboon shyness is not an aspect of the system which the park managers or humans living near the park can control. However, as the shyness parameter modulates the amount of risk the baboons are willing to take while crop-raiding, it can also be interpreted as a measure of how far the baboons are willing to travel to reach the crops. If the baboons are very shy they would, for instance, be less willing to travel a long distance to crop-raid. This provides an additional type of crop control -- planting attractive crops further from park edges. Thus, we can interpret Figure \ref{fig:huntalpvsgam} with this new framework, allowing us to consider how three of the potential management strategies interact -- less nutritious crops, planting further from the boundary, and human hunting. For example, if the crops are of high nutritional value and human hunting is occurring at medium levels, planting further away from the park boundary could successfully counteract the high proportion of time spent crop-raiding due to loss of apex predators. The requirement for both human hunting of baboons and planting further from the boundary in this example emphasizes the need to utilize multiple management strategies to control crop-raiding when apex predators are absent, which is not necessary when apex predators are present. At least two out of the three management strategies of less nutritional crops, planting further from the boundary, and human hunting must occur in order for management of crop-raiding to be successful if apex predators have been extirpated.

This paper highlights the strong positive effects of top-down control by apex predators on mesopredators. Mesopredator release can have wide-ranging deleterious effects. In our example it leads to increasing levels of crop-raiding. In other systems, trophic downgrading, the removal of apex predators from an ecosystem, can have far-reaching effects on species further down the food chain such as extinction and loss of habitat \citep{estes2011trophic,estes2013predicting}, and hence it is of great concern. These indirect effects of apex predators may not be realized until after the apex predators have been extirpated. This work substantiates the principle that all trophic levels needs to be maintained to promote healthy and balanced ecosystems \citep{estes2011trophic,ripple2014status}. Further, doing so can have knock-on benefits for humans, and can compliment other management strategies. The model presented here, and similar models, can be valuable tools in evaluating the potential impacts of suites of management strategies and therefore inform approaches for managing protected areas. 

\section*{Acknowledgments}

Early versions of this work were developed while SJR was a Postdoctoral Associate at the National Center for Ecological Analysis and Synthesis, a Center funded by NSF (Grant \#EF-0553768), the University of California, Santa Barbara, and the State of California. We wish to thank Andr\'{e} de Roos for his help and input at the early stages of this work, especially in model development.

\appendix
\section{Demography Calculations}

Using a life-table analysis, we calculated the maximum demographic potential of baboons in an unrestricted resource environment, using parameters from the literature. Olive baboon ({\it Papio anubis}) females can live 25-30 years. It is suggested that baboon females reach sexual maturity at 4-5 years of age and begin reproduction 1-2 years later \citep{altmann_determinants_1988} with an interbirth interval of 1-2 years \citep{smuts1989reproduction} although age of maturity and reproduction in the wild has been shown to vary with demographic and environmental factors \citep{beehner2006ecology,charpentier2008age}. They have been recorded as reproducing as early as 5 and continuing until over 20 years of age \citep{smuts1989reproduction}, however, fecundity falls off quite dramatically after this point.

We tabulated data from \citet{strum1982variations} and \citet{altmann1977life} to estimate fecundity and $\lambda$, the rate of population growth, in a wild population to ascertain how this compared with unrestricted life-history calculated measures. 

We set up our life table with 30 yearly age classes ($x$), with a constant mortality effect of 1/30 (1/maximum lifespan). We seeded this population with a large number (1000), so that rates were not influenced by small number problems. We set the age of first reproduction at the minimum of 5 years and termed at 25 years, to maximize reproductive lifespan. We then established fecundity by assuming minimum interbirth interval (1 year), with 1 successful offspring per event (but we chose not to include twinning), and a sex ratio skewed fully female. This yields a maximum $\lambda$ of 1.204, or a growth rate of 20.4\% annually. 

There are many ways to explore the range of differences our assumptions impose on this estimate: for example, assuming parity sex ratio at birth, we find this drops to 14\% ($\lambda=1.14$), because we have effectively halved female fertility. 

Using the data from \citet{strum1982variations} and \citet{altmann1977life} we calculated $\lambda$ based on total population counts, and found a geometric mean estimate (for exponential growth) of 1.025, or 2.5\% annual growth. In terms of female fecundity, there was a raw fecundity (average annual reproduction) of 0.576. 

If we combine this fecundity with our life table, assuming an equal sex ratio at birth, we set $m_x$ to 0.29, reducing $\lambda$ to 1.10, or 10\% annual growth. In order to approach the measured $\lambda$, while maintaining assumptions of constant mortality, we must impose a $>10$\% mortality rate. Clearly the life-table assumption of constant mortality and fecundity rates during reproductive life-span are inappropriate to real baboon demography, but this life history modeling exercise shows the capacity of baboons to respond to resources demographically, as suggested by \citet{bercovitch1993dominance}, in absence of any restrictions.

\section{Further Analysis}

We expand on our results to show more details of how the four main parameters of interest affect the proportion of time that the baboons choose to crop-raid. This includes further information on Figure 5 in the main text and a sensitivity analysis of changes in our parameters on the proportion of time crop-raiding.

\subsection{Effect of apex predator removal on baboons and crop-raiding levels}

\begin{figure}[!ht]
\includegraphics[width = 16cm]{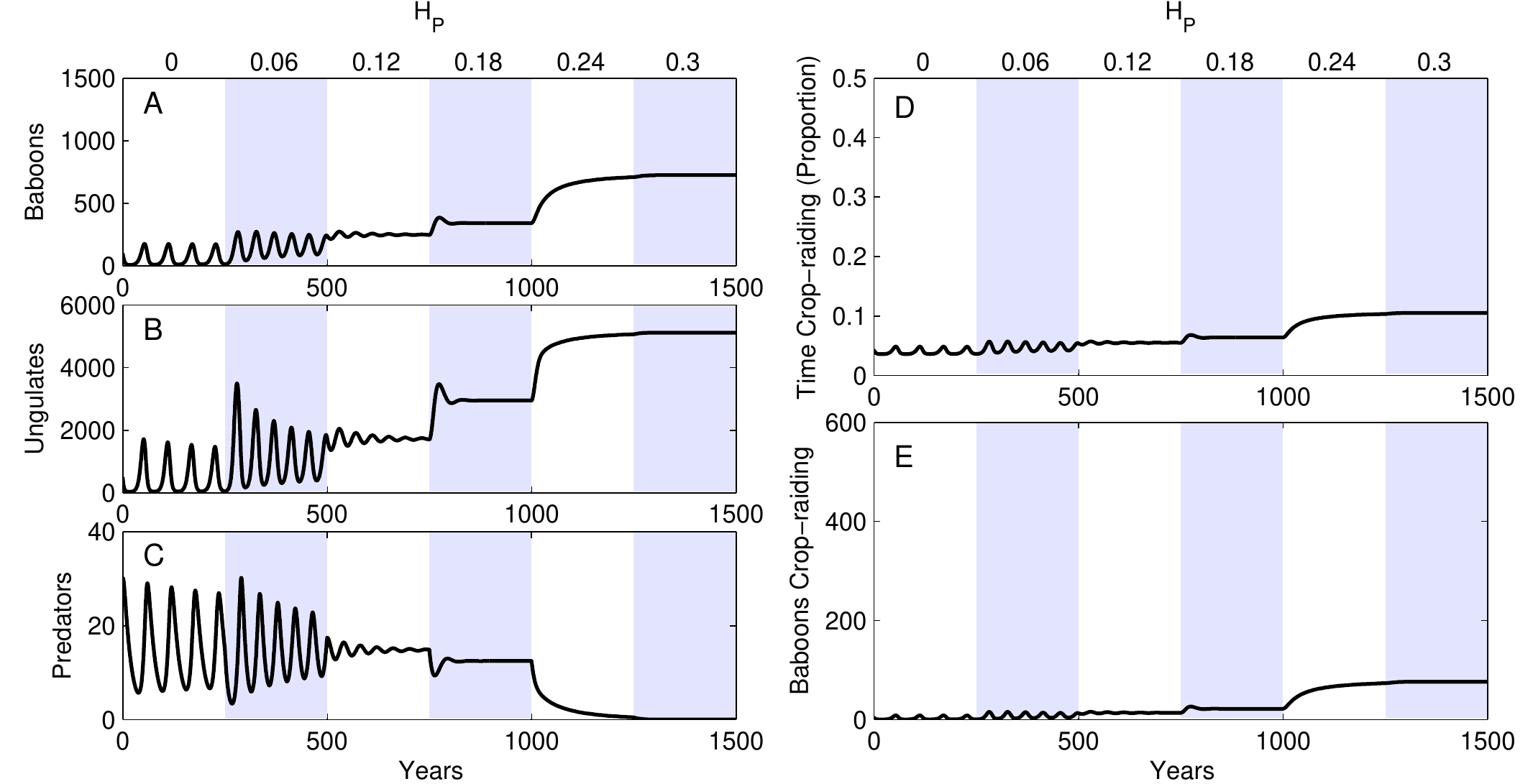}
\caption{Gradual increase of hunting on the apex predators ($H_P$) to show mesopredator release with loss of apex predators, alongside human hunting of baboons ($H_B$). In $A$, $B$ and $C$ the dynamics of baboons, ungulates and apex predators over time, respectively. In $D$ the proportion of time baboons choose to crop-raid, ($1-\tau$), as well as the potential damage caused when they do so in $E$, measured as the number of baboons raiding $(1-\tau)B$, are plotted against time. Parameter values are as in Table 1 in the main text apart from $H_B = 0.25$ and $H_P$ which varies from 0 to 0.3 every 250 years in increments of 0.06, as indicated on the top axis. The long time scale is used to show the cyclic dynamics at each level of hunting in more detail.} 
\label{app:GradPredHuntHB}
\end{figure}

In Figure 2 in the main text, we show how the loss of apex predators leads to a higher abundance of both baboons and ungulates, and this results in increases in the proportion of time that baboons crop-raid. Notably, there is a significant rise in the number of baboons that crop-raid, measured as $(1-\tau)B$. In Figure \ref{app:GradPredHuntHB}, we investigate the loss of apex predators on baboon abundance and proportion of time spent crop-raiding using the same technique as Figure 2 in the main text, but now with human hunting of baboons occurring.

From Figure \ref{app:GradPredHuntHB} we see that baboon abundance and proportion of time spent crop-raiding both increase as the apex predators are hunted to extinction. However, human hunting of baboons is now present which suppresses the baboon population and hence also the proportion of time they choose to crop-raid. Human hunting is successful at reducing the proportion of time crop-raiding because it is at a very high level, $H_B = 0.25$, which shows the potential for human hunting to be reasonably effective. However, even with hunting at its maximum level, it does not keep the proportion of time crop-raiding at the same low level as when apex predators are present.

\subsection{Mixed Hunting Strategy Expanded}

\begin{figure}[!ht]
\includegraphics[width = 16cm]{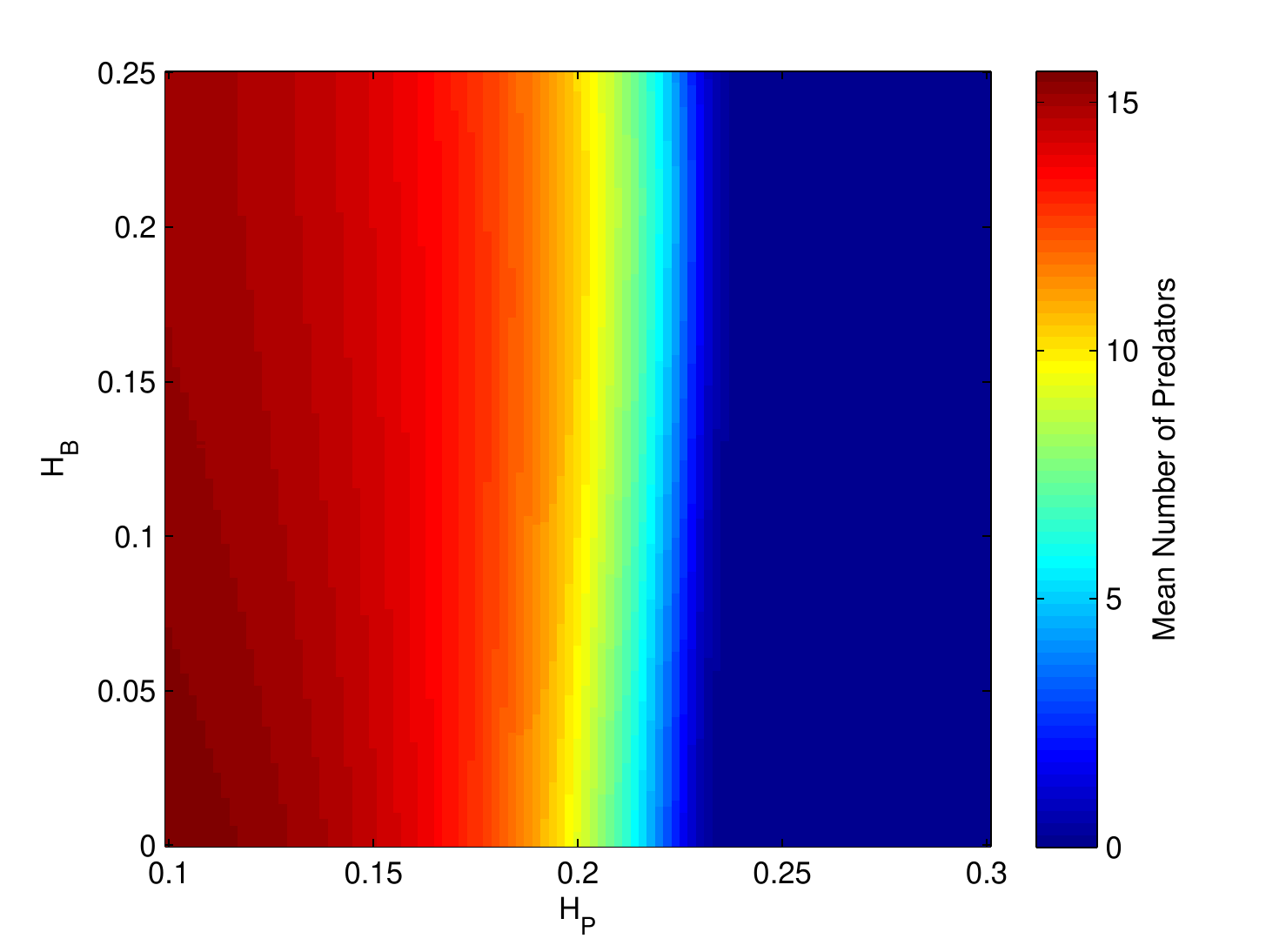}
\caption{The effect of human hunting of baboons ($H_B$) and apex predators ($H_P$) on the mean abundance of apex predators. All other parameters are as in Table 1 in the main text.} 
\label{app:huntpredabun}
\end{figure}

In Figure 5 in the main text, the proportion of time baboons crop-raid is shown against the combined strategies of hunting apex predators and baboons. This shows that it is possible to hunt apex predators up to $H_P = 0.23$ and still gain the beneficial apex predator suppression of baboons. However, it is useful to understand what effect this hunting of apex predators has on apex predator abundance. Thus for each value of $H_B$ and $H_P$ in Figure 5 in the main text, we calculate mean apex predator abundance over the last 100 years of the 500 year simulation (to allow the transient dynamics to fade), as shown in Figure \ref{app:huntpredabun}.

From Figure \ref{app:huntpredabun}, it is clear that $H_B$ does not have a significant effect on the number of apex predators as the color does not change in the vertical direction, apart from a slight curve in the middle region. This is because when there is high apex predator abundance baboons do not crop-raid. Hence changes in $H_B$ do not impact greatly on baboon abundance as hunting of baboons only occurs when they crop-raid. Therefore, if baboon numbers are not affected, there is no feedback to changes in apex predator abundance. However, for mid-range values of $H_P$, higher hunting of baboons will allow apex predators to survive a slightly higher level of hunting before extirpation. Figure \ref{app:huntpredabun} shows that lower levels of apex predator hunting will reduce the abundance of apex predators, although there will only be a significant decrease for $H_P > 0.16$. From Figure 5 in the main text, we know that as long as $H_P < 0.23$, it is possible to keep the proportion of time baboons crop-raid at low levels. Thus, there is a range of values for $H_P$ such that apex predator abundance can be reduced whilst at the same time maintaining the benefits of baboon suppression.

\subsection{Sensitivity Analysis}

\begin{figure}[!ht]
\includegraphics[width = 10cm]{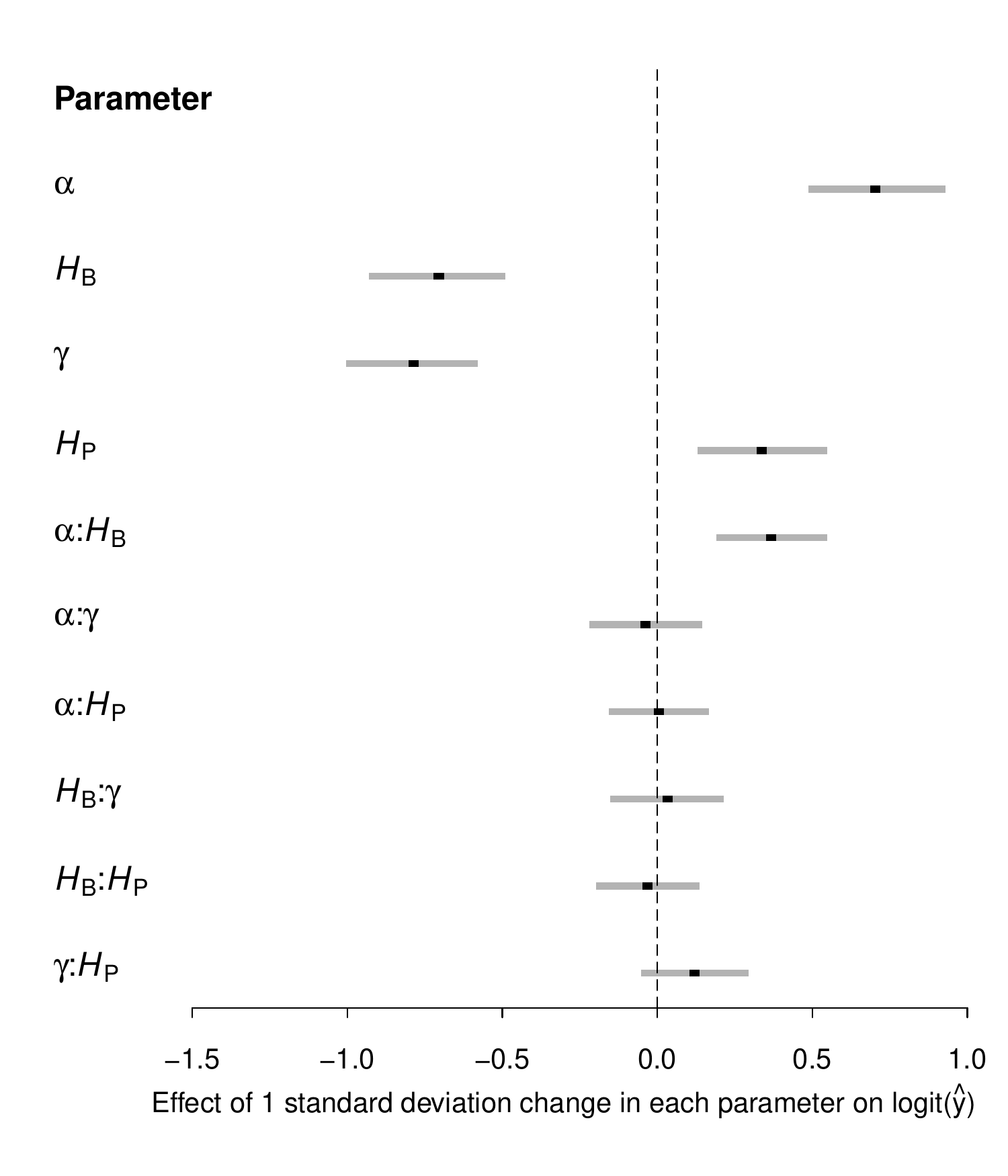}
\caption{A global sensitivity analysis showing the effect of one standard deviation change in each of the 4 parameters ($\alpha$, $\gamma$, $H_B$ and $H_P$) on the proportion of time spent crop-raiding. The black dot shows the value of the slopes and the interactions between slopes, with the grey bars indicating 95\% confidence intervals.} 
\label{app:sensitivity}
\end{figure}

In the main text, we focus on comparing one or two factors together that lead to changes in the proportion of time that baboons crop-raid. In Figure \ref{app:sensitivity}, we wish to compare the 4 main factors, and the two-way interactions between the factors. To do this, we followed the procedure in \cite{coutts2014meta} by using a generalized linear model to run a global sensitivity analysis of the proportion of time spent crop-raiding to changes in each of the four parameters of interest ($H_B$, $H_P$, $\alpha$ and $\gamma$). Briefly, the method involves running 3000 simulations of the model while each of the four parameters can take any value within a range. We used the same ranges as in the main text, i.e.\ $0<H_B<0.25$, $0 < H_P < 0.3$, $0.5<\alpha<1.8$ and $0<\gamma<1.2$, and used Latin hypercube sampling method in order to get the most representative sets of parameters for the 3000 simulations. For each simulation we calculated the average proportion of time spent crop-raiding, $1-\tau$, over the last 200 out of 1000 years. After standardizing the parameter inputs so that they can be compared more easily, we then ran a generalized linear model in R using the data. As the data is in terms of proportions, we use the binomial family which results in the logit link function.

From Figure \ref{app:sensitivity}, we can see that a one standard deviation increase in parameters $\alpha$ and $H_P$ leads to increases in the proportion of time spent crop-raiding, while a one standard deviation increase in $H_B$ and $\gamma$ would lead to decreases in time spent crop-raiding. These results are in agreement with Figures 2--5 in the main text. If crops are more nutritional (higher $\alpha$), baboons are encouraged out to crop-raid, as well as when their population is not being suppressed by apex predators (higher hunting of apex predators, $H_P$). Conversely, human hunting of baboons ($H_B$) and higher shyness of baboons ($\gamma$) will discourage them from venturing out of the park. These effects are all significant; however, the only two-way interaction significantly different from zero is that of $H_B$ and $\alpha$. Figure \ref{app:sensitivity} indicates that both $\alpha$ and $\gamma$ are the most important to change in order to reduce the proportion of time spent crop-raiding. Whilst it is not possible to change the shyness of baboons, planting crops further from the boundary can act as a proxy to increase the risks a baboon would face in order to crop-raid. Thus, planting less nutritional crops and planting them further from the boundary could be a very successful method to reduce crop-raiding. However, the benefits of this would have to be outweighed against the costs of different crop strategies and loss of usable farmland. The hunting of apex predators, $H_P$ is not shown to have as influential an impact as $\alpha$ in Figure \ref{app:sensitivity}. This is likely caused by the fact that $H_P$ has a threshold effect. Once $H_P$ is large enough the apex predators go extinct, thus further increases in $H_P$ will not affect the dynamics. Therefore comparison between the sensitivity of the model to changes in $H_P$ and the other parameters should be inferred with caution. From the main text, we know that the presence of apex predators significantly reduces the proportion of time spent crop-raiding.

\bibliographystyle{ecol_lett}

\end{document}